\documentclass[fleqn,usenatbib]{mnras}

\usepackage{newtxtext,newtxmath}

\usepackage[T1]{fontenc}

\DeclareRobustCommand{\VAN}[3]{#2}
\let\VANthebibliography\thebibliography
\def\thebibliography{\DeclareRobustCommand{\VAN}[3]{##3}\VANthebibliography}

\usepackage{graphicx}	
\usepackage{amsmath}	

\title[CD-30 Binary Evolution]{Modelling the AM CVn and Double Detonation Supernova Progenitor Binary System CD-30$^{\circ}$11223}

\author[Deshmukh et al.]{Kunal Deshmukh,$^{1,2}$\thanks{E-mail: astro.kunal.deshmukh@gmail.com}
Evan B. Bauer,$^{3}$
Thomas Kupfer,$^{4,2}$
and Matti Dorsch$^{5}$
\\
$^{1}$Institute of Astronomy, KU Leuven, Celestijnlaan 200D, 3001 Leuven, Belgium\\
$^{2}$Department of Physics and Astronomy, Texas Tech University, PO Box 41051, Lubbock, TX 79409, USA \\
$^{3}$Center for Astrophysics | Harvard \& Smithsonian, 60 Garden St, Cambridge, MA 02138, USA\\
$^{4}$Hamburger Sternwarte, University of Hamburg, Gojenbergsweg 112, 21029 Hamburg, Germany\\
$^{5}$Institut für Physik und Astronomie, Universit\"at Potsdam, 14476 Potsdam-Golm, Germany
}

\date{Accepted XXX. Received YYY; in original form ZZZ}

\pubyear{2023}

\begin{document}
\label{firstpage}
\pagerange{\pageref{firstpage}--\pageref{lastpage}}
\maketitle 

\begin{abstract}
We present a detailed modelling study of CD-30$^{\circ}$11223 (CD-30), a hot subdwarf (sdB)-white dwarf (WD) binary identified as a double detonation supernova progenitor, using the open-source stellar evolution software MESA. We focus on implementing binary evolution models carefully tuned to match the observed characteristics of the system including $\log g$ and $T_{\rm eff}$. For the first time, we account for the structure of the hydrogen envelope throughout the modelling, and find that the inclusion of element diffusion is important for matching the observed radius and temperature. We investigate the two sdB mass solutions (0.47 and 0.54 $M_{\odot}$) previously proposed for this system, strongly favouring the 0.47 $M_{\odot}$ solution. The WD cooling age is compared against the sdB age using our models, which suggest an sdB likely older than the WD, contrary to the standard assumption for compact sdB-WD binaries. Subsequently, we propose a possible alternate formation channel for CD-30. We also perform binary evolution modelling of the system to study various aspects such as mass transfer, orbital period evolution and luminosity evolution. Our models confirm CD-30 as a double detonation supernova progenitor, expected to explode $\approx55$ Myr from now. The WD accretes a $\approx0.17$ $M_{\odot}$ thick helium shell that causes a detonation, leaving a 0.30 $M_{\odot}$ sdB ejected at $\approx$750 km\,s$^{-1}$. The final 15 Myr of the system are characterised by helium accretion which dominates the system luminosity, possibly resembling an AM CVn-type system.

\end{abstract}




\section{Introduction}




Hot subdwarf B stars or sdBs are subluminous spectral type B stars that lie on or near the Extended Horizontal Branch on the Hertzsprung-Russel diagram. They are typically core-helium burning post-main sequence (MS) stars that are stripped of their hydrogen envelopes \citep{heb86,heb09,heb16}. In most cases, this is a consequence of a mass transfer phase with a binary companion \citep{nap04a,max01}. Moreover, studies show that binary interaction might be required for the formation of all sdBs \citep{pel20}. SdBs that result from an unstable mass transfer phase, a so-called common envelope ejection phase with a companion can end up in compact binary systems with periods less than a few days. The companion in such cases can be a MS star or a white dwarf (WD), and angular momentum loss via gravitational wave radiation can significantly shrink the orbit further \citep{han02,han03,nel10}.

In compact sdB-WD binaries, when the orbital period right after common envelope ejection is $\lesssim$2 hr, the sdB can overflow its Roche-lobe within its helium burning lifetime \citep{ibe87,ibe91,bau21}. Such systems are excellent candidates for helium accretion onto a white dwarf from an sdB donor. In this special case where the white dwarf is accreting helium-rich material from its companion, a so-called Double Detonation Supernova is possible \citep{ibe87,ibe91,liv90,liv95,fin10,woo11,wan12,she14,wan18,won23}. As the name suggests, such a supernova results from two detonations - first a helium shell detonation on the white dwarf surface, causing a second detonation inside the C/O core. Provided the conditions are favourable, the white dwarf can then explode as a thermonuclear supernova even at significantly sub-Chandrasekhar masses. It is also possible that the helium shell detonation does not cause a core detonation, and instead simply results in a faint .Ia Supernova followed by weaker helium flashes \citep{bil07,bro15}. Recently several studies presented evidence for transients consistent with a thick helium shell double detonation on a sub-Chandrasekhar-mass WD, leading to a peculiar type I supernova \citep{de19,de20, pol19, pol21, dong22, col22, pad23, gon23, liu23SN22joj}. 

So far only two systems are known that show such short orbital periods and sufficiently large masses for a potential supernova. The first known system was CD-30$^{\circ}$11223 \citep{ven12, gei13}. More recently \citet{kup22} found PTF1\,J2238+7430 to match the requirements for a double detonation progenitor. For PTF1\,J2238+7430 detailed modelling shows that the white dwarf formed after the sdB in the system. \citet{rui10} predicted that some fraction of compact sdB+WD binaries could exist where the sdB formed first in a phase of stable mass transfer and the white dwarf companion second during a common envelope phase.

CD-30$^{\circ}$11223 (CD-30 here onward) was first mentioned by \citet{ven12} and later identified as the first Double Detonation Supernova progenitor by \citet{gei13} (referred to as G13 here onward). G13 discovered CD-30 as part of the MUCHFUSS project \citep{gei11a} in a search of compact sdB binaries with massive companions. Along with a current orbital period of 70 minutes, the lightcurve analysis revealed ellipsoidal modulation of the sdB as well as eclipses of both components. Combining this with atmospheric properties obtained from spectra, the orbital parameters and component properties were constrained. Most notably, G13 presented two  possible solutions for the component masses with neither being preferred. Based on their results and binary evolution models, G13 predicted CD-30 to undergo a mass transfer phase and eventually undergo a thermonuclear runaway about 42 Myr from now. G13 modelled the helium accretion phase and assumed that the WD would explode after $0.1$\,$M_\odot$ of helium accretion. Here we present detailed modelling of CD-30 using the stellar evolution code MESA, including the hydrogen accretion phase and the impact of the possible mass, age and progenitor of the sdB. We also focus on the evolutionary modelling of the sdB to match the present-day observational constraints. In Sec.\,\ref{section:sed}, we provide updated parameter estimates for the system based on the {\it Gaia} parallax and improved constraints on the SED. Sec.\,\ref{section:sdB} presents detailed modelling of the sdB, and Sec.\,\ref{section:age} shows the impact of the age of the sdB and the WD companion. The binary evolution of CD-30 is discussed in Sec.\,\ref{section:binary}, and in Sec.\,\ref{section:conclusion} we finish with conclusions.

\section{Updated Parameters for the System}
\label{section:sed}

The atmospheric parameters for CD-30 reported in G13 were very typical for an sdB. The helium abundance $\log y$ = $-1.5$ was in broad agreement with the $\log y$ - $T_{\rm eff}$ trend followed by sdBs \citep{heb16}. In combination with lightcurve modelling, two possible parameter solutions for the sdB and WD were reported, particularly the mass combinations: $M_{\rm sdB}$ = 0.47~$M_\odot$, $M_{\rm WD}$ = 0.74~$M_\odot$ and $M_{\rm sdB}$ = 0.54~$M_\odot$, $M_{\rm WD}$ = 0.79~$M_\odot$ (with uncertainties of about 0.02 $M_\odot$ each). 

\begin{figure}
	\includegraphics[width=\columnwidth]{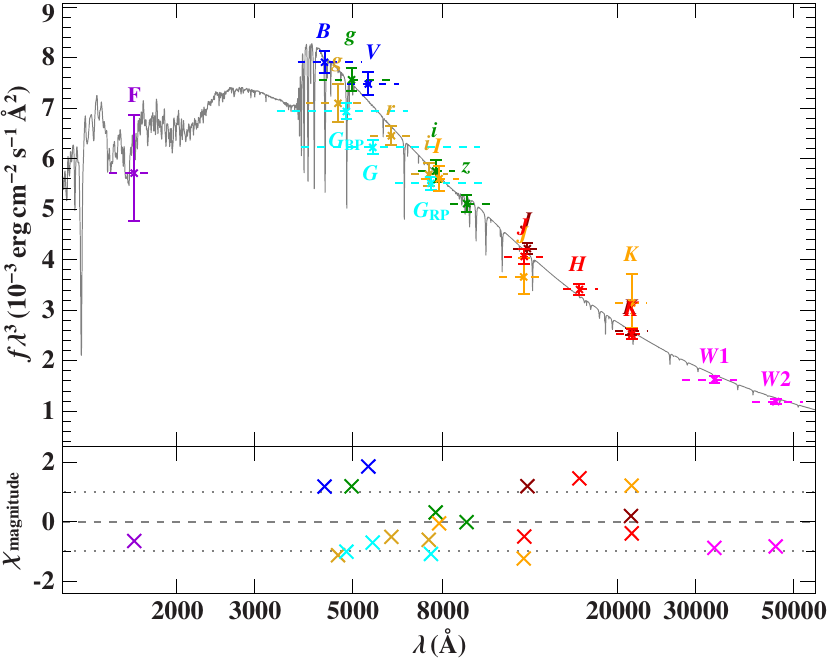}
    \caption{Spectral energy distribution fit for CD-30 based on 
    FAUST \citep[purple,][]{SED_FAUST}, 
    Johnson \citep[blue,][]{SED_APASS}, 
    SDSS \citep[green,][]{SED_APASS}, 
    SkyMapper \citep[yellow,][]{SED_Skymapper_DR2}, 
    \textit{Gaia} \citep[cyan,][]{SED_Gaia_EDR3}, 
    DENIS \citep[orange,][]{SED_DENIS}, 
    2MASS \citep[red,][]{SED_2MASS}, 
    VISTA/VHS \citep[dark red,][]{SED_VHS}, 
    and WISE \citep[pink,][]{SED_VHS} magnitudes. The best-fit model is shown in gray and error-weighted residuals are shown in the bottom panel. 
    }
    \label{SED_fit}
\end{figure}

Follow-up observations in X-rays with {\it XMM-Newton} were undertaken to investigate possible wind mass loss from the sdB surface, placing an upper limit of $\dot M_{\rm W} = 3\times10^{-13}$~$M_{\odot}\,\rm yr^{-1}$ \citep{mer14}. Updated astrometry from the {\it Gaia} mission allows measuring the mass of the sdB by combining the spectroscopic $T_{\rm eff}$ and $\log g$ with a spectral energy distribution (SED) fit, a method that is described in detail by \cite{Heber2018}. 
We constructed the SED of CD-30 (Figure\ \ref{SED_fit}) from archival photometry, ranging from the far-UV to the near-infrared. 
It is well reproduced by a synthetic spectrum computed for the atmospheric parameters of G13, as listed in Table~\ref{table:sed}. 
Free parameters in this SED fit were the angular diameter on the sky $\Theta$ and the color excess $E(44-55)$. 
The latter is caused by interstellar reddening, treated here using the empirical extinction curve of \cite{Fitzpatrick2019}; the best-fit reddening is low and consistent with the value of $0.04\pm 0.02$\,mag given by the ``Stilism'' 3D reddening map \citep{Capitanio2017}. 

For the computation of the stellar parameters, we used the \textit{Gaia} EDR3 parallax \citep{Gaia2021} with a corrected zero-point offset after \citet{Lindegren2021} and an inflated uncertainty according to equation 16 of \citet{El-Badry2021}. 
This combined with the angular diameter from the SED and the spectroscopic $T_{\text{eff}}$ and $\log g$ from G13 resulted in a mass of $0.47^{+0.07}_{-0.06}$\,$M_\odot$. 
This SED mass estimate excludes the high-mass solution ($0.54\pm0.02$\,$M_\odot$) at a formal 1-$\sigma$ confidence. 
%
%
In terms of radius, this difference is more pronounced due to smaller uncertainties: the SED fit ($0.167 \pm 0.005$\,$R_\odot$) and light curve analysis ($0.179 \pm 0.003$\,$R_\odot$) disagree at 1.5-$\sigma$ confidence. 
In contrast, the low-mass solution of G13 agrees almost perfectly with the stellar parameters derived by the SED method. 
%


\begin{table}
\caption{Measured parameters and results obtained from the SED fit. Spectroscopic inputs are from G13. Uncertainties are stated for 68\% confidence, using median Monte-Carlo values for the derived stellar parameters. 
}
\label{table:sed}
\renewcommand{\arraystretch}{1.15}
\begin{tabular}{lr}
\hline\hline
Parameter & Value \\
\hline
Effective temperature $T_{\mathrm{eff}}$ (prescribed) & $29200 \pm 400$\,K \\
Surface gravity $\log (g / \mathrm{cm\,s^{-2}})$ (prescribed) & $5.66 \pm 0.05$ \\
Helium abundance $\log(n(\textnormal{He})/n(\textnormal{H}))$ (prescribed) & $-1.5\pm 0.07
$ \\
Orbital Period & 70.53 minutes \\
\hline
Parallax $\varpi$  & $2.86 \pm 0.07$\,mas \\
Color excess $E(44-55)$ & $0.043 \pm 0.006$\,mag \\
Angular diameter $\log(\Theta\,\mathrm{(rad)})$ & $-10.666 \pm 0.006$ \\
\hline
Radius $R = \Theta/(2\varpi)$  & $0.167 \pm 0.005$\,$R_\odot$ \\
Mass $M = g R^2/G$ & $0.47^{+0.07}_{-0.06}$\,$M_\odot$ \\
Luminosity $L/L_\odot = (R/R_\odot)^2(T_\mathrm{eff}/T_{\mathrm{eff},\odot})^4$  & $18.3^{+1.5}_{-1.4}$ \\
\hline
\end{tabular}
\end{table}

\section{Modelling the sdB}
\label{section:sdB}

Previous modelling of CD-30 has approximated the masses of the system components by taking the average of the two parameter solutions obtained observationally \citep{gei13,bro15,bau17}. More recently, modelling compact sdB-WD systems in close accordance with observed atmospheric parameters has proved to be effective at yielding constraints that have implications on their formation and evolution \citep{kup22}. Furthermore, such modelling can also be used to investigate the two solutions from G13 in light of the updated parameters from Section \ref{section:sed}. This provides strong motivation to model CD-30 in detail, especially owing to it being a double detonation supernova progenitor. 

The low helium abundance relative to solar observed at the surface of the sdB in CD-30 suggests that it is important to account for some degree of sedimentation and atmospheric stratification when modelling its surface, which is also relevant for matching the observed radius and $\log g$. In general, element diffusion in sdB atmospheres is considered crucial due to their depleted surface abundances \citep{saf94}. \citet{byr18} modelled sdB progenitors to study the role of diffusion up to sdB formation. Quick depletion of helium and heavier elements was seen in their models, even beyond observed constraints on surface abundances, and additional physics was deemed necessary for accurate modelling. The hydrogen envelope which consists mainly of hydrogen and helium can be considerably affected by diffusion of helium. Moreover, the effects can directly show up in the $T_{\text{eff}}$ - $\log g$ evolutionary tracks. In this work, we included element diffusion in a similar fashion as \citet{byr18} with a qualitative focus on its role in determining $T_{\text{eff}}$ - $\log g$ evolutionary tracks, although a few models without diffusion were also explored for comparison.

We used the release version r22.05.1 of the MESA Stellar Evolution codes \citep{Paxton2011,Paxton2013,Paxton2015,Paxton2018,Paxton2019,Jermyn2023} to model the sdB as a single star in this section, and in a binary in Section \ref{section:binary}. All MESA model files in this work are available at \doi{10.5281/zenodo.10022986}. The MESA EOS is a blend of the OPAL \citep{Rogers2002}, SCVH \citep{Saumon1995}, FreeEOS \citep{Irwin2004}, HELM \citep{Timmes2000}, PC \citep{Potekhin2010}, and Skye \citep{Jermyn2021} EOSes. Radiative opacities are primarily from OPAL \citep{Iglesias1993,Iglesias1996}, with low-temperature data from \citet{Ferguson2005} and the high-temperature, Compton-scattering dominated regime by \citet{Poutanen2017}. Electron conduction opacities are from \citet{Cassisi2007} and \citet{Blouin2020}. Nuclear reaction rates are from JINA REACLIB \citep{Cyburt2010}, NACRE \citep{Angulo1999} and additional tabulated weak reaction rates \citep{Fuller1985, Oda1994,Langanke2000}. Screening is included via the prescription of \citet{Chugunov2007}. Thermal neutrino loss rates are from \citet{Itoh1996}.

Studying the sdB evolutionary tracks on the $T_{\text{eff}}$ - $\log g$ diagram is useful to test models in reference to observed parameters. Our goal was to investigate the two solutions from G13. We used MESA to create sdB models following the procedure from \citet{bau21}. This was done by evolving an MS star until the start of core helium burning, followed by implementing artificially enhanced winds to quickly remove the envelope until a specified total mass of hydrogen remained at the surface of the star.

The sdB envelope mass is typically $\lesssim 0.02$\,$M_\odot$ \citep{heb86} rendering it too thin to sustain hydrogen shell burning. As such, there are no direct observational constraints on the envelope mass, which motivated us to explore a wide range, provided there is no hydrogen shell burning. SdBs descendent from $\lesssim$2.0 $M_{\odot}$ MS stars have envelopes that are similar in composition to their progenitors. On the other hand, sdBs descendent from higher mass progenitors can possess envelopes enriched in helium and hence denser due to having undergone partial nuclear processing in the former convective core on the main sequence. We followed the approach taken by \citet{bau21} to use the total hydrogen mass ($M_{\rm H}$) in the sdB instead of the envelope mass ($M_{\rm env}$) as a variable parameter.

We used results from \citet{bau21} to estimate the mass of the MS progenitor for a given sdB mass. SdBs with masses within $0.47\pm0.03$ $M_{\odot}$ and $0.54\pm0.02$ $M_{\odot}$ as reported by G13 were considered. The sdB-MS progenitor mass relation is sensitive to parameters such as the progenitor metallicity and convective overshooting efficiency to a small extent \citep{ost21}. However, in this work we assume solar metallicity and no convective overshooting. The models therefore produce a representative set of sdBs across the relevant range of masses, but the relation between MS progenitor mass and final sdB mass may be somewhat imprecise.

The sdBs obtained were evolved through the core helium burning phase to produce $T_{\rm eff}$ - $\log g$ evolutionary tracks. We used the predictive mixing scheme \citep{Paxton2018} to model the convective helium core. This enabled the helium core to grow with time to masses close to asteroseismology predictions (e.g., \citealt{VanGrootel2010a,VanGrootel2010b,Charpinet2011,Charpinet2019,ost21}). It also ensured longer sdB lifetimes and avoided "breathing pulses" due to repeated division of the convective core that is likely a numerical artifact \citep{Paxton2019}. Refer to \citet{ost21} for a detailed discussion on the modelling of sdB convective cores. 

To investigate the role of element diffusion, we modelled each sdB with and without diffusion. In the former case, we included diffusion in the envelope region, while keeping it off in the core where it is not expected to affect our models. In the latter case, it was off throughout.  

In the following subsections, the two sdB mass solutions presented by G13 are discussed in detail.

\subsection{0.47 $M_\odot$ sdB}

The $\approx$0.47 $M_\odot$ canonical sdB mass has a broad range of possible MS progenitor masses. Stars with initial masses in the range $0.8 - 2.0$ $M_\odot$ are not able to start helium burning non-degenerately and require an off-center helium flash to ignite helium. This typically happens when the core mass reaches about 0.47 $M_\odot$ (with a slight $\approx 0.01\,M_\odot$ dependence on metallicity \citep{ost21}), making it the so-called canonical mass. On the other hand, for more massive stars with initial masses $\gtrsim 2.3$ $M_\odot$, helium burning can start non-degenerately. The core masses at the beginning of helium burning in this case can range from about 0.3 $M_\odot$ upwards, increasing monotonically with the progenitor mass. Consequently, a canonical mass sdB can originate from two fundamentally different types of progenitors, as discussed below.

\subsubsection{High mass progenitor}
\label{subsubsection:high}

The monotonic relation between progenitor mass and sdB mass for stars $\gtrsim 2.3$ $M_\odot$ allows for a small range of progenitor masses that can produce a 0.47 $M_\odot$ sdB. For the assumptions that we made in our MESA models, a progenitor mass in the range of $3.60-3.75$ $M_\odot$ results in our desired sdB.

For these progenitors, the envelope contains partially burned hydrogen owing to the receding convective core during their MS evolution. The core never becomes degenerate and its boundary with the envelope is not as sharp as the degenerate case. Consequently, the fraction of hydrogen is less than 70\%, the rest being mostly helium, and the envelope is compact.

\begin{figure}
	\includegraphics[width=\columnwidth]{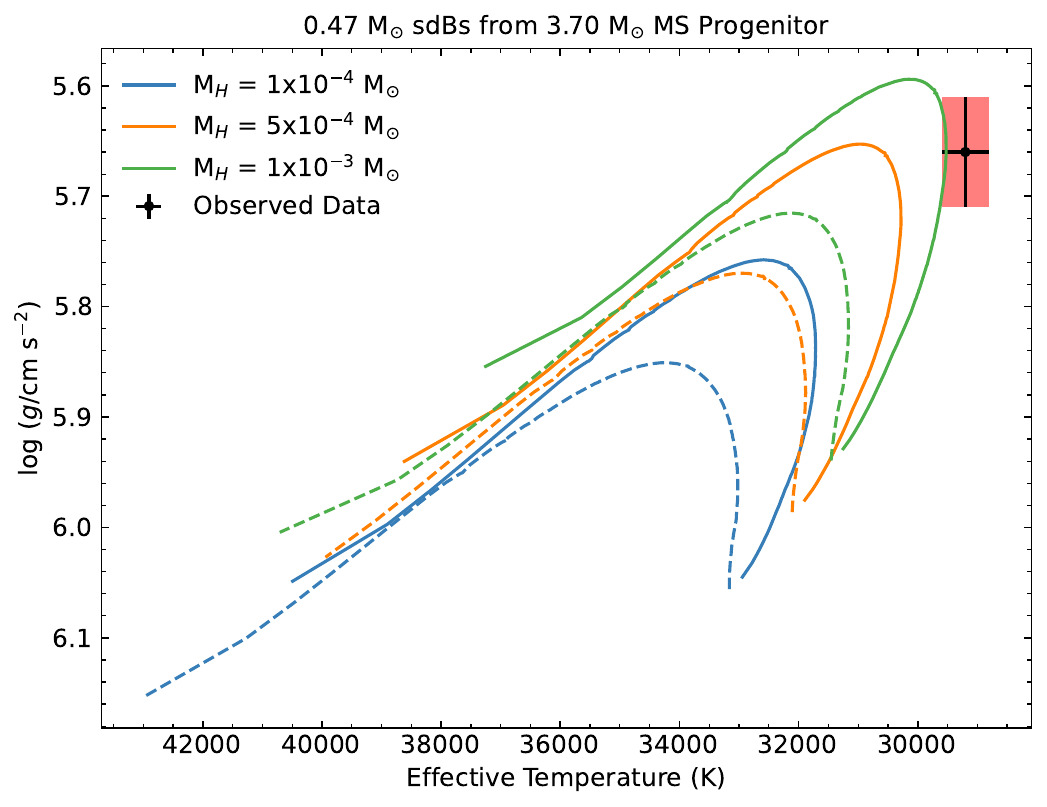}
    \caption{$T_{\text{eff}}$ - $\log g$ evolutionary tracks for three sdB models derived from a 3.70 $M_\odot$ MS progenitor with $M_{\rm H}$ = $1\times10^{-4}$, $5\times10^{-4}$ and $1\times10^{-3}$ $M_\odot$, along with the observed data and corresponding 1-$\sigma$ error region. The dashed lines show tracks without diffusion whereas the solid lines show tracks with diffusion.}
    \label{fig:him1}
\end{figure}

Figure\,\ref{fig:him1} shows the evolutionary tracks for sdB models derived from a 3.70 $M_\odot$ progenitor with $M_{\rm H}$ values $1\times10^{-4}$, $5\times10^{-4}$ and $1\times10^{-3}$ $M_\odot$ with and without element diffusion. As mentioned earlier, $M_{\rm H}$ is the total mass of hydrogen contained in the envelope. It is worth noting that the total envelope mass is larger than this value since it also contains a significant amount of helium.

In the tracks shown in Figure\,\ref{fig:him1}, there appears to be a shift towards lower $\log g$ and effective temperatures with increasing hydrogen mass. Additionally, the presence of diffusion also led to a similar trend. Since the envelopes for sdBs coming from high mass progenitors are rich in helium, diffusion could cause the helium to sink away from the surface efficiently to make the envelope more inflated and move the tracks towards lower $\log g$ and $T_{\text{eff}}$ values. 

\begin{figure}
	\includegraphics[width=\columnwidth]{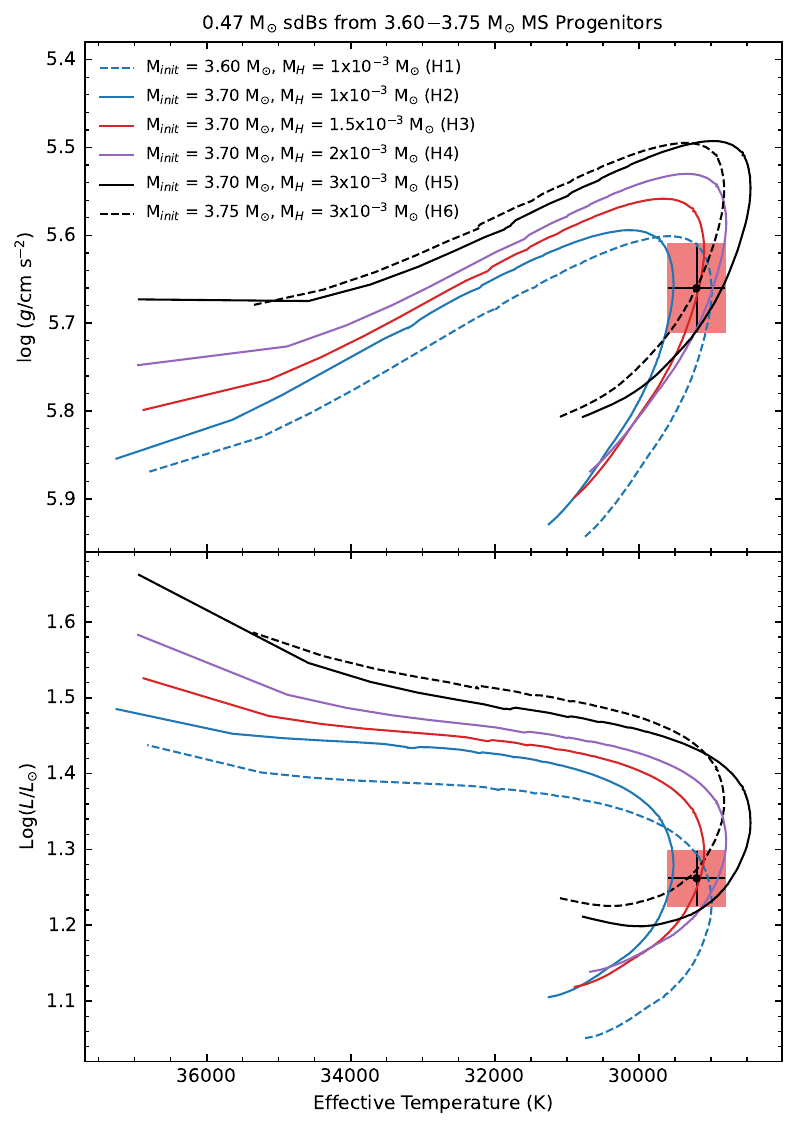}
    \caption{Top panel: $T_{\text{eff}}$ - $\log g$ evolutionary tracks for four sdB models derived from a 3.70 $M_\odot$ MS progenitor with $M_{\rm H}$ = $1\times10^{-3}$, $1.5\times10^{-3}$, $2\times10^{-3}$ and $3\times10^{-3}$ $M_\odot$ that pass through the red region, along with two models from 3.60 $M_\odot$ and 3.75 $M_\odot$ progenitors with  $M_{\rm H}$ = $1\times10^{-3}$ and $3\times10^{-3}$ respectively; Bottom panel: Same tracks in the $T_{\text{eff}}$ - $\log L$ space with the corresponding observed data and error region.}
    \label{fig:him2}
\end{figure}

To get the best models consistent with the error bars of observations, the $M_{\rm H}$ parameter was fine tuned to make the tracks pass through the red box. We adopted the physically motivated approach of including diffusion in all these models. Figure \ref{fig:him2} shows four possible models that were considered from a 3.70 $M_\odot$ MS progenitor, with $M_{\rm H}$ = $1\times10^{-3}$, $1.5\times10^{-3}$, $2\times10^{-3}$ and $3\times10^{-3}$ $M_{\odot}$, named as H2, H3, H4 and H5 respectively.  Additionally, the progenitor masses in the neighborhood of 3.70 $M_\odot$ were also able to yield suitable sdB tracks with slightly different $M_{\rm H}$ values. Two such models are shown in Figure \ref{fig:him2} for progenitor masses M$_{\rm init}$ = 3.60 and 3.75 $M_\odot$ with $M_{\rm H}$ = $1\times10^{-3}$ and $3\times10^{-3}$ $M_\odot$, named as H1 and H6 respectively. The top panel shows evolutionary tracks in the $T_{\rm eff}$ - $\log g$ space whereas the lower panel shows them in the  $T_{\rm eff}$ - $\log L$ space.

An important result from Section \ref{section:sed} is the well constrained luminosity of the sdB based on the measured $T_{\rm eff}$ and an excellent {\it Gaia} parallax for CD-30. We take advantage of this measurement by using the $T_{\rm eff}$ - $\log L$ space and modelling the sdBs to satisfy this additional constraint as well. It is interesting to note that in Figure \ref{fig:him2}, although both sets of tracks simultaneously pass through the red region, the error bars on the $\log L$ parameter correspond to a shorter evolutionary time and are hence better suited for age considerations as will be discussed in Section \ref{section:age}.

Since the tracks start at considerably different points and represent a range of progenitor and envelope masses, all models were considered for further analysis.

\subsubsection{Low mass progenitor}
\label{subsubsection:low}

For 0.8-2.0 $M_\odot$ MS stars, the core mass required for the helium flash is around 0.47 $M_\odot$. We considered progenitors in the range 1.0-1.9 $M_\odot$. All sdBs obtained were nearly identical as is expected due to the common denominator of a helium flash. Progenitors with masses more than 1.9 $M_\odot$ gradually transition into the non-degenerate helium burning regime and the sdB masses drop.

Since the sdB models obtained from stripping most of the envelope of a MS star in the 1.0-1.9 $M_\odot$ range are similar, any of those sdB models in principle would be appropriate as a representative model. In accordance with this, we chose a 1.80 $M_\odot$ progenitor motivated by its relatively shorter lifetime, which will turn out to be the most plausible in future discussion sections.

For stars that undergo a helium flash, the core and the envelope are separated by a relatively sharp boundary. The envelope essentially has a composition similar to the initial composition of the star, which in this case would be solar composition. Therefore, the thin sdB envelope is expected to be about 70\% hydrogen. 

\begin{figure}
	\includegraphics[width=\columnwidth]{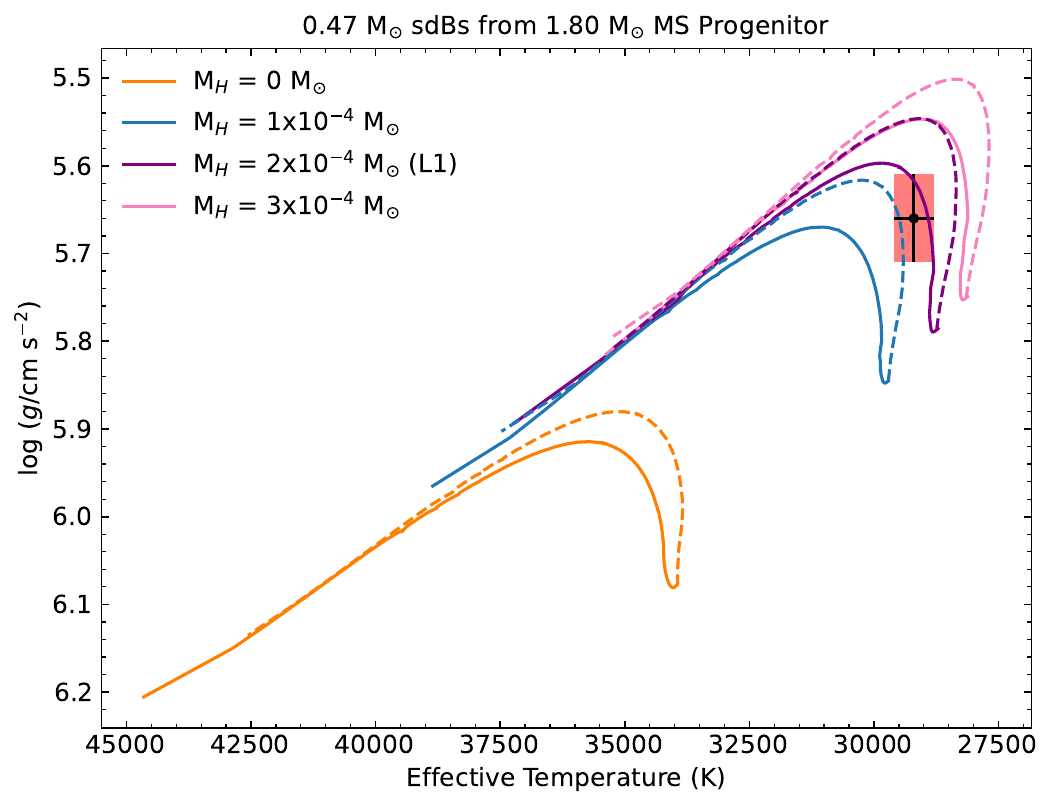}
    \caption{$T_{\text{eff}}$ - $\log g$ evolutionary tracks for three sdB models derived from a 1.80 $M_\odot$ MS progenitor with $M_{\rm H}$ = 0, $1\times10^{-4}$, $2\times10^{-4}$ and $3\times10^{-4}$ $M_\odot$. The dashed lines show tracks without diffusion whereas the solid lines show tracks with diffusion.}
    \label{fig:lom}
\end{figure}

Figure \ref{fig:lom} shows the evolutionary tracks on the $T_{\text{eff}}$ - $\log g$ diagram for sdB models derived from a 1.80 $M_\odot$ progenitor with $M_{\rm H}$ values of 0, $1\times10^{-4}$, $2\times10^{-4}$ and $3\times10^{-4}$ $M_\odot$. The evolutionary tracks were very sensitive to the hydrogen mass, shifting towards considerably lower values of $T_{\text{eff}}$ and $\log g$ even with increments of the order of $1\times10^{-4}$ M$_\odot$. 

The evolutionary tracks in the absence and presence of diffusion were seen to be similar, with a general trend of slightly lower $T_{\text{eff}}$ and $\log g$ values for models without diffusion. For these models, in contrast to the models of the previous subsection, the most important effect of diffusion is that it removes metals from the surface layers, changing the opacity and leading to a slightly more compact structure. This can be seen most clearly in the model with zero hydrogen envelope, where diffusion still leads to a similar change in the $T_{\text{eff}}$ - $\log g$ track due to removing metals from the layers near the photosphere. The effect of diffusion is smaller for these models than it is for the high-mass progenitor models. Keeping this in mind, the final model was chosen with diffusion enabled. The sdB model with $M_{\rm H}$ = $2\times10^{-4}$ $M_\odot$ was found to be the best fit to observed data, as shown in Figure \ref{fig:lom}. However, it is important to note that there was a small range of envelope masses that led to largely equivalent tracks passing through the error bar region, and any of those models would be appropriate for further evolution. The $M_{\rm H}$ = $2\times10^{-4}$ $M_\odot$ model, named as L1, was therefore chosen as the representative model for the low mass progenitor case.

In conclusion of this subsection, we found 0.47 $M_\odot$ sdB models satisfying the observed atmospheric parameters with possible origins from both low and high mass progenitors. There was no conclusive evidence to differentiate between the two broad possibilities and both were deemed suitable for binary evolution modelling.

\subsection{0.54 $M_\odot$ sdB}

\begin{figure}
	\includegraphics[width=\columnwidth]{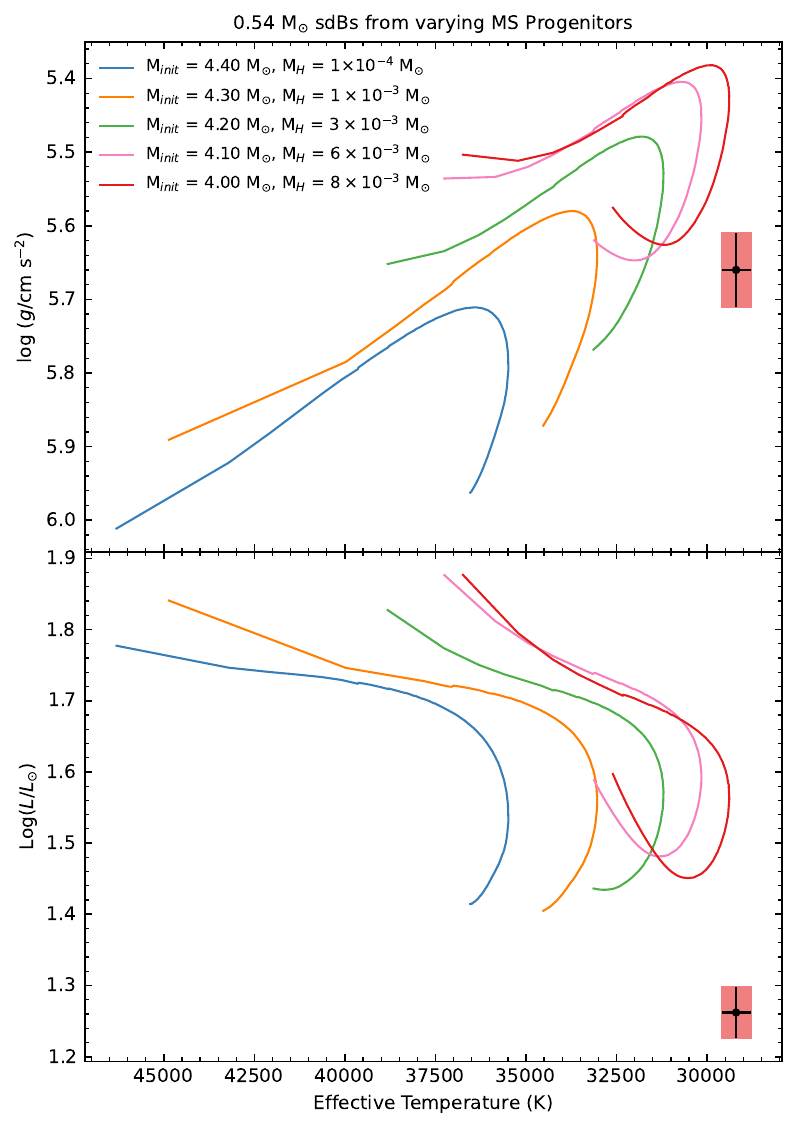}
    \caption{$T_{\text{eff}}$ - $\log g$ (top panel) and $T_{\text{eff}}$ - $\log L$ (bottom panel) evolutionary tracks for 0.54 $M_{_\odot}$ sdBs derived from varying MS progenitor masses and corresponding envelope masses. All models have diffusion enabled. Both plots strongly disfavour the 0.54 $M_{_\odot}$ sdB solution.}
    \label{fig:him4}
\end{figure}

Presented as the second possible solution for the sdB in CD-30 by G13, 0.54 $M_\odot$ sdB models were put to a similar test as above. Unlike the 0.47 $M_\odot$ case, a 0.54 $M_\odot$ sdB can only be derived from a high mass progenitor. Following the approach described in Section \ref{subsubsection:high}, we explored a wide range of progenitor masses and $M_{\rm H}$ values. Figure \ref{fig:him4} shows evolutionary tracks for sdB models derived from progenitors in the range 4.0-4.4 $M_\odot$ and $M_{\rm H}$ values in the range $1\times10^{-3}$ to $8\times10^{-3}$ $M_\odot$. The combinations of MS progenitor mass and the $M_{\rm H}$ parameter are chosen such that the sdB mass is constant at 0.54 $M_{\odot}$. Envelope diffusion is enabled in all models in this case, although it is worth noting that the effect of diffusion on the evolutionary tracks here is expected to be similar to Section \ref{subsubsection:high}.

The evolutionary tracks for models derived from 4.2-4.4 $M_\odot$ MS progenitors are similar to those in Section \ref{subsubsection:high}. However, towards the lower end of MS masses (paired with the higher end of $M_{\rm H}$ values), the tracks show a different behaviour for the initial few Myr, rising in both $T_{\text{eff}}$ as well as $\log g$. This can be attributed to the significant residual hydrogen shell burning due to higher amounts of hydrogen present in the envelope. The sdB tracks from 4.00 and 4.10 $M_\odot$ progenitors are shown as representative models for this case and progenitors below 4.00 $M_\odot$ are not considered due to the increasing role of hydrogen shell burning.

Based on the tracks shown in the top panel of Figure \ref{fig:him4}, none of the 0.54 $M_\odot$ sdB models pass through the 1-$\sigma$ uncertainty region of the observed $T_{\text{eff}}$ - $\log g$ values. In particular, the models derived from 4.2-4.4 $M_\odot$ progenitors that have negligible hydrogen shell burning are all a few thousand Kelvin hotter than the observed $T_{\text{eff}}$. Additionally, the bottom panel shows an even stronger inconsistency, most notably in the high value of luminosity in all models regardless of progenitor or envelope masses. These more luminous models with higher core masses clearly disfavour the 0.54 $M_\odot$ solution for the sdB in CD-30, as supported by the SED fitting results from Section \ref{section:sed} as well. It is therefore not considered further for the binary evolution modelling of CD-30.

\section{Age of the system}\label{section:age}

Compact sdB-WD binaries are thought to form via the common envelope ejection channel. Although the phenomenon of common envelope ejection is ubiquitous in all kinds of compact binaries, its inherent three-dimensional nature makes it extremely difficult to model numerically. With our current knowledge of this phenomenon, it is not possible to trace back a compact binary system to its pre-common envelope properties such as its period and component masses, making the true age of the system uncertain.

In the context of this work, the age of the system was defined as the time passed since the most recent common envelope ejection. To find the age of CD-30, a simple approach was followed – finding the ages of the sdB and WD and taking the smaller number of the two. 

In non-eclipsing ellipsoidal sdB-WD systems, obtaining the temperature and radius of the WD is typically not possible because it is too faint to observe. For CD-30 however, the eclipses enabled G13 to determine the temperature of the WD to be around 24,700 K. Based on Section \ref{section:sdB}, the 0.47 $M_\odot$ sdB solution is favored, which was paired with a 0.74 $M_\odot$ WD. Combining the knowledge of the temperature and mass of the WD, we employed the WD cooling age tables provided by \citet{bed20} to estimate the age of the WD in CD-30. We found the cooling age to be around 39 Myr with an uncertainty of 10 Myr. 

Similar to the WD age, the sdB age was defined as the time passed since its formation, with formation defined as the beginning of helium core burning, which coincides with the stripping of the envelope to form the sdB in our construction. We then timed the closest approach of the evolutionary tracks to the reference $T_{\text{eff}}$ - $\log L$ which was discussed in Section \ref{subsubsection:high} to obtain the sdB age. Table \ref{tab:age} summarizes the sdB ages for different sdB models that were considered.

\begin{table}
	\centering
	\caption{Current age of the sdB in CD-30 for a range of possible MS progenitor masses ($M_{\rm init}$) and envelope hydrogen masses $M_{\rm H}$}
	\label{tab:age}
	\begin{tabular}{|c|c|c|c|c|} 
		\hline
		Model & $M_{\rm init}$ ($M_{\odot}$) & $M_{\rm H}$ ($M_{\odot}$) & SdB Age (Myr) \\
 		\hline
		L1 (0.46 $M_{\odot}$) & 1.80 & 2$\times 10^{-4}$ & 84 $\pm$ 18 \\
		H1 (0.47 $M_{\odot}$) & 3.60 & 1$\times 10^{-3}$ & 118 $\pm$ 16 \\
            H2 (0.46 $M_{\odot}$) & 3.70 & 1$\times 10^{-3}$ & 84 $\pm$ 17 \\
            H3 (0.46 $M_{\odot}$) & 3.70 & 1.5$\times 10^{-3}$ & 78 $\pm$ 18 \\
            H4 (0.46 $M_{\odot}$) & 3.70 & 2$\times 10^{-3}$ & 70 $\pm$ 17 \\
            H5 (0.47 $M_{\odot}$) & 3.70 & 3$\times 10^{-3}$ & 57 $\pm$ 20 \\
            H6 (0.48 $M_{\odot}$) & 3.75 & 3$\times 10^{-3}$ & 39 $\pm$ 19 \\
		\hline
	\end{tabular}
\end{table}

The uncertainties in sdB age were calculated as the amount of time the tracks spend within the 1-$\sigma$ error bars (red regions in the evolutionary track plots) of the observations. The L1 model gave a representative value for the sdB age coming from a wide range of low mass progenitors since all such sdBs closely resemble each other. The models H1 to H6 showed a significant variation in the sdB age and could have varying implications on the origin of the system, as discussed in the following paragraphs.

We find two categories of models: those for which the sdB age is clearly older than the WD (L1,H1,H2,H3 and H4), and those for which the ages are comparable, with the possibility that the sdB is younger given the error bars (H5 and H6).

Classically, it has been assumed that compact sdB-WD systems originate from the sdB progenitor undergoing a common envelope ejection phase with its WD companion \citep{han03,gei11a}. It is implicit in this scenario that the WD has already formed and ejects the envelope of its companion which forms the sdB. The support for this assumption from observations is scarce since determining the WD age depends heavily on its temperature, which in turn can only be constrained in high-inclination eclipsing systems. 

Previous modelling of CD-30 by G13 (see also \citealt{bro15,bau17}) was also based on the classical formation channel, assuming that the WD formed first, followed by the sdB. However, a recent study of a compact sdB-WD binary by \citet{kup22} revealed an sdB older than the WD for the first time, making it necessary to explore an additional formation channel.

\begin{figure*}
	\includegraphics[scale=0.26]{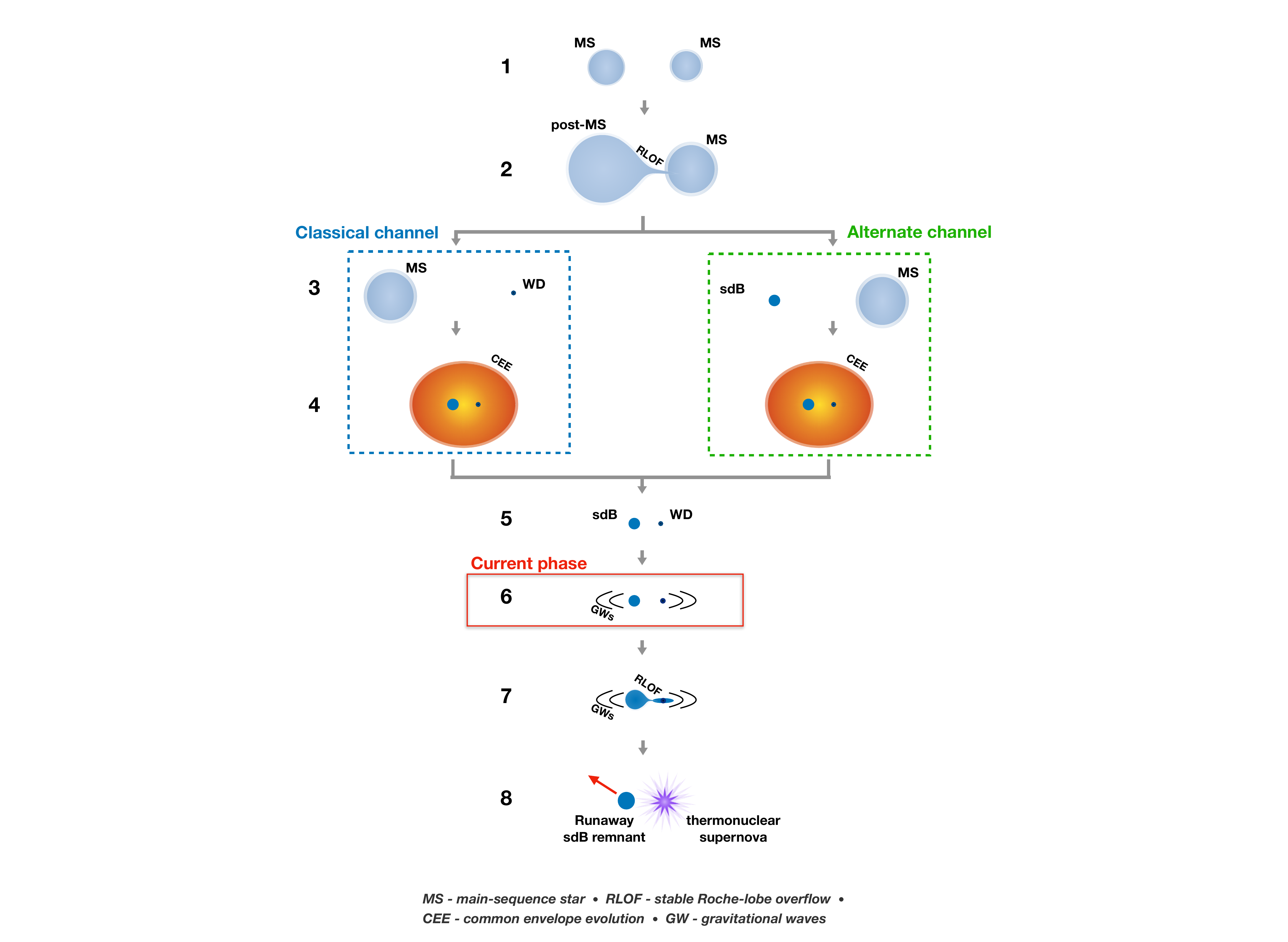}
    \caption{An illustration of two possible evolutionary pathways for CD-30 based on the sdB being younger or older than the WD. The channel on the left represents the so-called classical pathway of the WD forming first, followed by the sdB resulting from common envelope ejection. The channel on the right, labelled as the ``alternate" channel, represents a pathway with the order of formation reversed, leading to the sdB being older than the WD.}
    \label{fig:channels}
\end{figure*}

As described in \citet{kup22}, and shown here in Figure\,\ref{fig:channels} as the alternate channel, the distinguishing feature of this younger WD channel is the formation of the sdB first via stable mass transfer, followed by the formation of the WD via common envelope ejection. The resulting system is a compact sdB-WD binary with the sdB age more than the WD age. This channel closely resembles the scenario described as ``typical detached evolution'' for forming double C/O WDs discussed in section 3.1 of \citet{rui10}. The only difference in the context of forming a system such as CD-30 is that the post-common envelope period would need to be slightly shorter so that the system can come into contact within the helium-burning lifetime of the sdB, rather than burning out to form a detached double WD system.

In the context of CD-30, the family of possible solutions for the sdB in our MESA models is inconclusive as to whether it is older or younger than the WD. An sdB older than the WD would make CD-30 one of only two such known systems. On the other hand, an sdB younger than the WD would simply support the classical channel as discussed earlier. The modelling done in this work was therefore indicative of two possibilities for the formation of the system and did not favor one over the other. 


For the L1 solution, the younger WD channel is preferred based on the models created in this work. Qualitatively, this can be realised by considering two MS stars of similar masses (say 1.8 $M_{\odot}$ and 1.7 $M_{\odot}$) of which, the primary star evolves first and transfers mass to its companion via stable Roche-lobe overflow at the tip of the Red Giant Branch to form the sdB. An additional consequence of the stable mass transfer is the widening of the orbit. This is followed by the quick evolution of the now-massive secondary star to enter its Asymptotic Giant Branch (AGB) phase and overflow its Roche-lobe. Given the extreme mass ratio, the system undergoes a common envelope ejection phase to form a compact sdB-WD system that we see as CD-30 now. This scenario, although qualitative, encouraged the sdB progenitor mass choice from Section \ref{subsubsection:low}. The sdB in CD-30 could equally well be modelled from progenitors in the 1.0-1.9 $M_{\odot}$ range as indicated by sdB models. However, since the secondary star needs to evolve into a 0.74 C/O $M_{\odot}$ WD, a higher total mass of the system is preferred. 

For the solutions H1-H4, a similar approach as L1 could be followed. However, assuming the initial masses to be around 3.70 $M_{\odot}$ and $\approx$3 $M_{\odot}$ implies that the system needs to lose a much higher amount of mass to evolve into CD-30. For such a system, assuming the mass ratio allows stable Roche-lobe overflow, the primary evolves into the sdB via stable mass transfer of its envelope to the secondary. Subsequently, the secondary evolves into its giant phase and undergoes a common envelope ejection to leave behind a compact sdB-WD system that is CD-30. This scenario faces the major challenge of matching evolutionary timelines of the two components. In particular, while the primary forms the sdB and is currently in the core helium burning phase, the secondary has to accrete mass, burn out the helium in its core to form a 0.74 $M_{\odot}$ C/O core and also undergo a common envelope ejection in its AGB phase to form the WD. The relative timescales for the evolution of the two stars would need to be somewhat fine-tuned for this scenario to work out, but it is worth noting that binary population synthesis has realized such a scenario requiring the same sequencing in at least some cases \citep{rui10}. While it is difficult to make quantitative estimates for the component lifetimes without detailed modeling for the progenitor binary evolution in this scenario, it is plausible that the He core for the star that forms the currently observed $\approx0.74\,M_\odot$ WD would be sufficiently massive to evolve quickly enough to overtake its companion. Because the core mass--luminosity relation for He core burning stars is steep, the He core mass would only need to be $\approx$20\% larger than the $0.47\,M_\odot$ mass of the sdB companion to evolve roughly twice as fast and reach the WD cooling sequence while its sdB companion still has 10s of Myr left in its 165~Myr core-burning lifetime.

Finally, the solutions H5 and H6 do not provide a preferred older/younger component in CD-30. While the same argument as H1-H4 would be plausible here, the classical channel can also explain the formation of CD-30. As shown in Figure\,\ref{fig:channels}, this channel is characterised by the formation of the WD before the sdB. The primary evolves first to form the WD. The secondary then evolves into a giant and undergoes the common envelope ejection phase to form the sdB in a compact binary with the WD. 
Although the sequence of steps may appear less fine-tuned in the classical channel, it is worth noting that the young ages of both binary components in this system still require a surprisingly narrow window for both stars to evolve off the main sequence in quick succession.

The age of the youngest component in the binary system also allows us to estimate the post-common envelope orbital period of the system. This is essentially the period at which it exited common envelope based on the inspiral time up to the current point given the two masses. For L1 and H1-H4, the younger WD age gives a post-common envelope period of the system to be around 88 minutes. For H5-H6, it could be as short as 80 minutes.

To summarise the formation scenarios, the multiple sdB solutions shown in Table \ref{tab:age} indicate a broad range of possibilities. A common factor among all scenarios is the significant uncertainty introduced by common envelope evolution as an intermediate step between the main sequence phase and the current phase of CD-30. Additional uncertainties introduced by the sdB as well as WD models also need to be factored in. The scenarios discussed above are therefore only qualitative attempts at telling the full story for CD-30. Rigorous modelling to test each of those scenarios is beyond the scope of this work.
In the next section, future evolution of CD-30 as a binary is discussed.

\section{Binary Evolution}
\label{section:binary}

\begin{figure*}
	\includegraphics[scale=0.8]{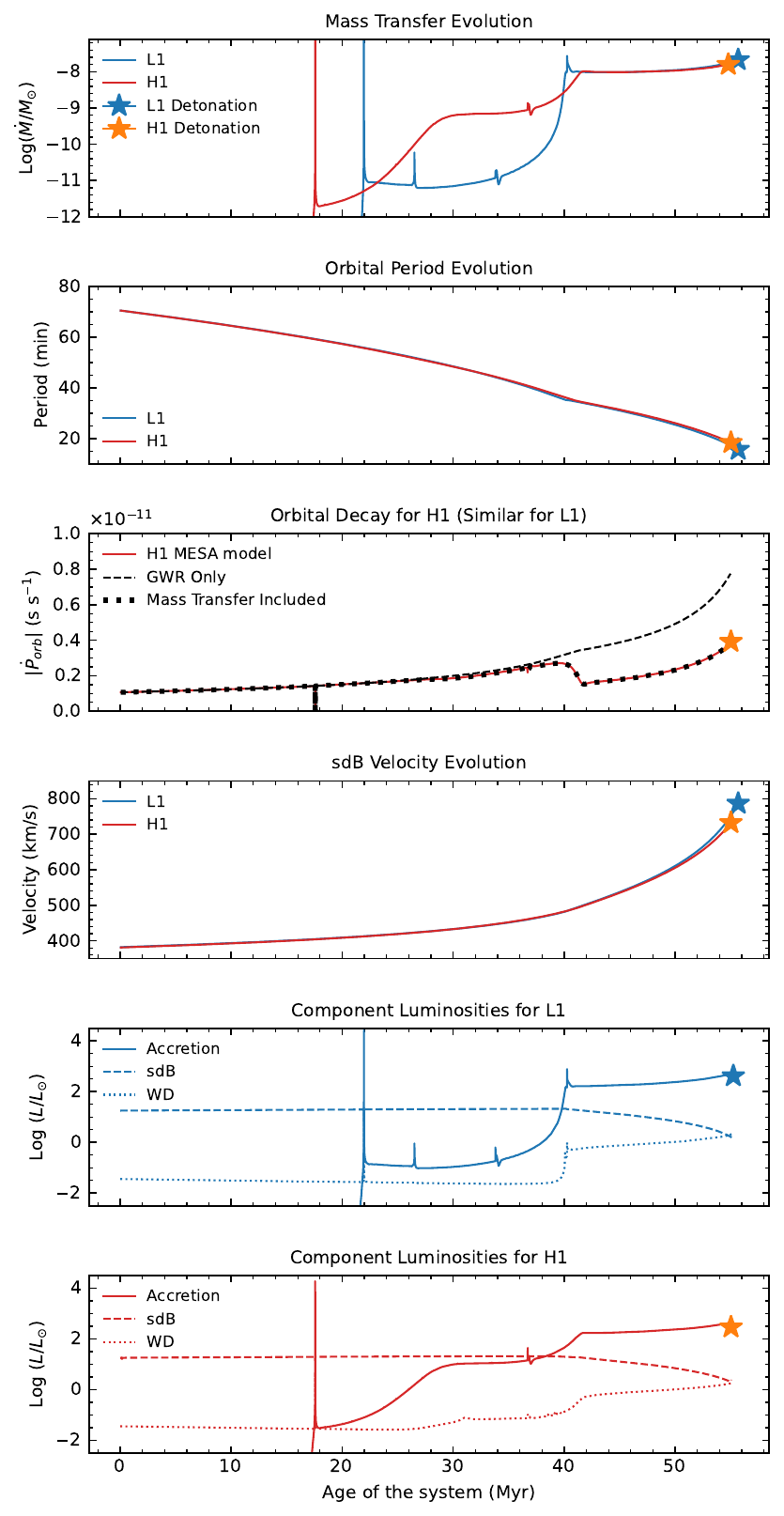}
    \caption{Results from the binary evolution of CD-30 from present day till detonation with MESA for L1 and H1 sdB models, shown in blue and red respectively. The panels are described in detail in Section \ref{section:binary}.}
    \label{fig:bin}
\end{figure*}

The current 70-minute orbital period of CD-30 is expected to shrink gradually due to angular momentum loss via gravitational wave radiation until the sdB eventually overflows its Roche-lobe and begins mass transfer to the WD.

We employed the MESA binary evolution module to evolve the sdB and WD in CD-30 together. Roche-lobe radii in the binary system were computed using the fit of \citet{Eggleton1983}.  Mass transfer rates in the Roche-lobe overflowing phase were determined following the prescription of \citet{Ritter1988}. We assumed fully conservative mass transfer from the sdB to the WD. All angular momentum losses from the system are assumed to be due to gravitational wave radiation, with 
\begin{equation}
    \dot J_{\rm gr} = - \frac{32}{5c^{5}} \left(\frac{2\pi G}{P_{\rm orb}}\right)^{7/3} \frac{(M_{1}M_{2})^{2}}{(M_{1}+M_{2})^{2/3}}~.
\end{equation}

The current day sdB models were obtained as per Section \ref{section:sdB} by setting the stopping condition for evolutionary tracks as the closest approach to the observed $T_{\text{eff}}$ - $\log L$ values. We evolved a 0.74 $M_\odot$ C/O white dwarf model to cool to the observed temperature of 24,700 K.  All sdB models from Table \ref{tab:age} in combination with the WD were considered to investigate the mass transfer, orbital period evolution and eventual thermonuclear runaway on the WD surface. The models H1-H6 were nearly equivalent and therefore we chose H1 as the representative model for sdBs derived from high mass progenitors. Accordingly, models L1 and H1 were considered for Figure \ref{fig:bin} which summarizes the most important features of the binary evolution of CD-30.

As the system evolves from present day and the orbit shrinks due to gravitational wave radiation, the sdB fills its Roche-lobe for both L1 and H1 at about 20 Myr. This marks the onset of stable mass transfer of the hydrogen envelope to the WD  as seen in the top panel of Figure \ref{fig:bin}. Although the envelope of sdBs is only a small fraction of the total mass, it forms a considerable portion of the physical size owing to its low density. Consequently, it takes about 20 Myr more for the sdB to transfer its envelope. As discussed earlier, the composition of hydrogen envelopes for L1 and H1 was significantly different, the former being about 70\% hydrogen and the latter being much less than that. Consequently, this mass transfer phase is also significantly different for the two cases, with the L1 model having a lower rate due to a sparser envelope occupying the same geometrical space.

The glitches in the mass transfer rate for both models during hydrogen mass transfer around 34 Myr(L1) and 37 Myr(H1) were caused due to a limitation of the tabulated input physics for the opacity as the surface of the donor becomes more hydrogen depleted. We verified that this glitch occurred at a hydrogen mass fraction of precisely X=0.1, which corresponds to a grid point in the composition grid for opacity and EOS tables. The change in opacity interpolation as the model evolved across this grid point led to a temporary change in radius that caused the mass transfer rate to be briefly discontinuous, but the model then converged back to the previous mass transfer rate, and this did not affect the subsequent evolution.

It is known that accumulation of hydrogen on WDs can lead to nova outbursts, which are computationally expensive and non-trivial to model with MESA (e.g. \citealt{wolf13,bau21}). In principle, it is possible to include these novae as part of our binary evolution model, as in \cite{bau21}. However, for the purpose of modeling CD-30 as a double detonation supernova progenitor, the hydrogen novae do not play a significant role in the final outcome since the key factor is the later accretion of helium-rich material. Consequently, the novae were artificially turned off on the WD surface during the envelope mass transfer phase. We accomplish this by setting energy production by nuclear burning to zero in the MESA WD accretor model as long as any hydrogen remains in the WD. There is therefore no instability as the transferred hydrogen burns away to helium as it is compressed and heated underneath newly accreted material. Eventually the hydrogen envelope of the donor is fully transferred, and once the underlying helium begins to transfer, the last of the hydrogen in the WD quickly burns away. We then turn full nuclear burning back on so that the later thermonuclear instability of the accreted helium envelope can indicate when a detonation is likely to occur.

At around 40 Myr, the envelope is exhausted in both cases and helium accretion takes off at a higher mass transfer rate of about $10^{-8}$~$M_{\odot}\,\rm yr^{-1}$. After accreting $\approx$0.17 $M_\odot$ helium-rich material for about 15 Myr, the thick helium shell accumulated undergoes a thermonuclear runaway at its base caused by the $^{14}{\rm N}(e^-, \nu) {^{14}{\rm C}} ( \alpha, \gamma) {^{18}{\rm O}}$ (NCO) reaction chain triggering a $3\alpha$ runaway \citep{bau17}. Similar to Section 5.1 from \citet{bau21}, the critical density for a detonation of about $10^{6}$ g\,cm$^{-3}$ \citep{woo94, woo11, neu17} is comfortably surpassed at the ignition location on the WD, where it is $\approx1.7\times10^{6}$ g\,cm$^{-3}$. Such detonations of thick He shells are expected to transition to the C/O core causing a second detonation, resulting in an explosion of the WD \citep{pol19,she21}, however with different spectroscopic signatures than ``normal'' Type Ia supernovae \citep{de19,liu23,pad23,gon23,liu23}.  

The binary models in this work therefore establish CD-30 as a double detonation supernova progenitor. It is important to note that all sdB models therein had helium burning lifetimes longer than the timescale of binary evolution of CD-30. We can quantify this timescale approximately by calculating the merger time based simply on gravitational wave radiation, which is given by 
\begin{equation}
    \tau_{\rm merge} = \frac{5}{256} \frac{c^{5}(M_{1}+M_{2})^{1/3}P_{\rm orb}^{8/3}}{(4\pi^{2})^{4/3}G^{5/3}M_{1}M_{2}}~.
\end{equation}
Using the current system parameters $M_{1}$ = 0.47 $M_{\odot}$, $M_{2}$ = 0.74 $M_{\odot}$ and $P_{\rm orb}$ = 70 min, we obtain the timescale to be around 46 Myr, significantly shorter than the sdB helium burning lifetimes of our models. This was an important factor in the fate of CD-30 as a double detonation supernova, since an earlier end to helium burning might have evolved the system into a double WD binary. For CD-30 however, all sdB models lead to the same result from binary evolution.

The second and third panels in Figure \ref{fig:bin} show the evolution of the orbital period and orbital period decay for CD-30 respectively. The period of the system is about 20 minutes at the time of explosion. The orbital period decay essentially follows the $\dot P_{\rm orb}$ due to gravitational wave radiation until the onset of helium mass transfer. Once the helium mass transfer is underway, there is an additional significant $\dot P_{\rm orb}$ term from the changing mass ratio that makes the overall $\dot P_{\rm orb}$ slower. For conservative mass transfer with total mass transfer rate $\dot M$, and neglecting any spin of the stars in the binary, $\dot P_{\rm orb}$ should evolve according to (e.g. \citealt{bur23}):
\begin{equation}
    \frac{\dot P_{\rm orb}}{P_{\rm orb}} = 3\frac{\dot J_{\rm orb}}{J_{\rm orb}} + 3\frac{\dot M}{M_{\rm donor}}(1-M_{\rm donor}/M_{\rm accretor})~.
\end{equation}
Looking at the third panel with this equation and the mass transfer evolution in mind for model H1, the first $\approx20$ Myr are consistent with angular momentum loss only by GWR. The envelope transfer phase causes a slight deviation owing to a small $\dot M$ term, followed by a significant deviation with the onset of helium accretion at a much higher $\dot M$. $\dot P_{\rm orb}$ evolution both with and without the $\dot M$ term are shown for comparison.

The fourth panel shows the sdB velocity evolution for CD-30. The velocity at the time of explosion represents the terminal velocity of the sdB as a remnant runaway star. Both L1 and H1 models have runaway velocities around 750 km\,s$^{-1}$, as expected for remnants from He star donors to double detonation supernovae \citep{bau19,neu21,neu22}.

The last two panels show bolometric luminosities of the sdB, WD and accretion components of CD-30 for L1 and H1 sdB models respectively. The sdB and WD luminosities are taken directly from the MESA models, whereas the accretion luminosity is estimated as the gravitational potential energy lost in the accretion disk $L_{\rm acc} \approx GM_{\rm WD} \dot M/R_{\rm WD}$. The system luminosity is dominated by the sdB before the onset of mass transfer, and also through the envelope mass transfer in the L1 case. However, for both L1 and H1, the onset of helium mass transfer marks a change in the dominant source of bolometric luminosity. The accretion luminosity exceeds that of the sdB by over an order of magnitude during this phase. From an observational perspective, this implies that the light we receive from CD-30 could be dominated by the helium-rich accretion disk in at least some wavelength bands. The resulting implication is a likely AM CVn-like appearance of CD-30 during this phase, which lasts for about 15 Myr (similar to the ``He-star'' donor model class for AM CVns discussed in e.g., \citealt{yun08,nel10,bau21}).

Furthermore, for the H1 case in the last panel, there is a $\approx10$ Myr long phase from around 28 to 38 Myr where the sdB and accretion luminosities are comparable. The duration of this phase is significant and corresponds to the transfer of the helium-rich part of the envelope. Observationally, although it is non-trivial to predict how such a phase would look like, our models support the possible existence of a new class of accreting sdB-WDs showing significant flux from both the accretion disk and donor star. Ongoing and upcoming photometric and spectroscopic surveys can perhaps shed some light on these systems.

\section{Conclusions}
\label{section:conclusion}

CD-30 is a compact sdB-WD binary first identified as a double detonation supernova progenitor by \citet{gei13} (G13). They employed photometric and spectroscopic methods to determine the orbital and atmospheric parameters of the the system. They also presented two possible solutions for the component properties - particularly the component mass combinations of $M_{\rm sdB}$ = 0.47 $M_\odot$, $M_{\rm WD}$ = 0.74 $M_\odot$ and $M_{\rm sdB}$ = 0.54 $M_\odot$, $M_{\rm WD}$ = 0.79 $M_\odot$. Recent observations from {\it Gaia} provide a precise parallax for CD-30. Subsequently, we obtained a good constraint on its luminosity, along with other improved parameters.

With the primary motivation to investigate CD-30 from a theoretical perspective and compare it to observational results, we created MESA models for the sdB in CD-30. We followed an approach similar to \citet{bau21} when creating our sdB models. The low surface abundance of helium in CD-30, which is also the case for many sdBs, necessitates the inclusion of diffusion when modelling such systems. We implemented diffusion in the envelope of our models and also compared them to models without diffusion for a sanity check. The observed parameters used for reference were $T_{\rm eff}$, $\log g$ and $\log L$. Consequently, we used $T_{\rm eff}$ - $\log g$ and $T_{\rm eff}$ - $\log L$ parameter spaces to study the evolutionary tracks of our models.

The two possible sdB masses presented by G13 were taken into consideration. Based on the relation between the sdB mass and its MS progenitor mass, a range of viable progenitor masses was explored. Furthermore, a wide range of envelope masses was also explored, since it is not constrained by any observable properties. For the 0.47 $M_{\odot}$ sdB, we derived sdB models from 1.80 $M_{\odot}$ (low mass) and 3.60 - 3.75 $M_{\odot}$ (high mass) MS progenitors with varying envelope masses that were consistent with observations. A notable difference between the sdBs derived from low mass and high mass progenitors was the structure of the envelope, the former being sparser and the latter being denser and richer in helium. 

For the 0.54 $M_{\odot}$ sdB, we used many combinations of MS progenitor mass and envelope mass to obtain the desired mass for our models. These combinations covered a broad range of 4.0 - 4.4 $M_{\odot}$ progenitor masses, which were all inconsistent with observations. Particularly, the precise measurement of $\log L$ from {\it Gaia} showed an even stronger inconsistency with our 0.54 $M_{\odot}$ sdB models. We therefore strongly favour the 0.47 $M_{\odot}$ sdB solution and considered our consistent models for further steps.

CD-30 is one of very few sdB-WD systems that are at high enough inclination to show both eclipses in its lightcurve. This enabled G13 to determine the WD temperature, which we used to estimate a cooling age of 39 $\pm 10$ Myr for the WD. The sdB evolutionary tracks enabled us to find the current age of the sdB by comparing them to the observed data. The sdB ages were in a broad range from 39-118 Myr.

The widely accepted channel for compact sdB binary formation is characterised by the MS progenitor going into a common envelope with the companion to form an sdB, thus making the sdB the younger component of the two. The sdB ages obtained from our MESA models broadly fall into two categories - one where the sdB is older and one where the sdB and WD are of comparable ages. The older sdB (or younger WD) case calls for a different formation channel where the sdB forms first and its companion goes into common envelope to form the WD. This would make CD-30 one of only two known systems so far that indicate this alternative scenario.

With a set of plausible 0.47 $M_{\odot}$ sdB models and a 0.74 $M_{\odot}$ WD model cooled to its current temperature, we modelled CD-30 with the MESA binary code. In summary, the results of the binary evolution were - (a) the orbit shrinks due to gravitational wave radiation for about 20 Myr, when the sdB Roche-lobe is filled; (b) the Roche-lobe overflow leads to the onset of envelope mass transfer from the sdB to WD for about 20 Myr; (c) after the envelope is exhausted, helium accretion begins at a higher mass transfer rate and about 15 Myr later, there is a thermonuclear runaway on the WD; (d) the total time until double detonation is about 55 Myr; (e) at the time of explosion, the sdB mass is 0.30 $M_{\odot}$ and the WD mass is 0.91 $M_{\odot}$, $\approx$0.17 $M_{\odot}$ of which is the thick helium shell; (f) the sdB is thrown away at $\approx$750 km\,s$^{-1}$, which is consistent with runaway stars; (g) the bolometric luminosity calculated for accretion exceeds that of the sdB in the helium accretion phase, possibly leading to an AM CVn-like appearance.

\section*{Acknowledgements}

We thank Stephan Geier for useful discussion about the sdB parameters for CD-30. TK acknowledges support from the National Science Foundation through grant AST \#2107982, from NASA through grant 80NSSC22K0338 and from STScI through grant HST-GO-16659.002-A. Co-funded by the European Union (ERC, CompactBINARIES, 101078773). Views and opinions expressed are however those of the author(s) only and do not necessarily reflect those of the European Union or the European Research Council. Neither the European Union nor the granting authority can be held responsible for them.

\section*{Data Availability}

All photometric data used for the spectral energy distribution fit are publicly available. 
All MESA input files and work directories for producing the simulations in this work are publicly available at \doi{10.5281/zenodo.10022986}.



\bibliographystyle{mnras}
\bibliography{mesa} 








\bsp	
\label{lastpage}
\end{document}